%% file: 0main.tex
\pdfoutput=1

\documentclass[11pt]{article}

\usepackage[preprint]{acl}

\usepackage{times}
\usepackage{latexsym}

\usepackage[T1]{fontenc}

\usepackage[utf8]{inputenc}

\usepackage{microtype}

\usepackage{inconsolata}

\usepackage{graphicx}
\usepackage{booktabs}
\usepackage{enumitem}
\usepackage{multirow}
\usepackage{algorithm}
\usepackage{algpseudocode}
\usepackage{listings}
\usepackage{listings}
\definecolor{codegreen}{rgb}{0,0.6,0}
\definecolor{codegray}{rgb}{0.5,0.5,0.5}
\definecolor{codepurple}{rgb}{0.58,0,0.82}
\definecolor{backcolour}{rgb}{0.95,0.95,0.92}
\definecolor{deepblue}{rgb}{0,0,0.5}
\definecolor{deepred}{rgb}{0.6,0,0}
\definecolor{deepgreen}{rgb}{0,0.5,0}

\lstdefinestyle{mypython}{
    backgroundcolor=\color{backcolour},   
    commentstyle=\color{codegreen},
    keywordstyle=\color{magenta},
    emph={self},  
    emphstyle=\color{deepred},    %
    numberstyle=\tiny\color{deepblue},
    stringstyle=\color{codepurple},
    basicstyle=\footnotesize\ttfamily,
    breakatwhitespace=false,        
    xleftmargin=0pt,
    breaklines=true,                 
    captionpos=b,                    
    keepspaces=true,                 
    numbers=left,                    
    numbersep=3pt,                  
    showspaces=false,                
    showstringspaces=false,
    showtabs=false,                  
    tabsize=2,
    frame=none,                      %
    rulecolor=\color{backcolour},    %
    lineskip=-0.2em,                 %
    language=Python,
    morekeywords={with},             %
}

\definecolor{neuralbackcolour}{rgb}{0.9375, 0.96875, 0.9960}
\lstdefinestyle{neural}{
    backgroundcolor=\color{neuralbackcolour},   
    commentstyle=\color{codegreen},
    keywordstyle=\color{magenta},
    numberstyle=\tiny\color{deepblue},
    stringstyle=\color{codepurple},
    basicstyle=\footnotesize\ttfamily,
    breakatwhitespace=false,
    xleftmargin=0pt,
    breaklines=true,                 
    captionpos=b,                    
    keepspaces=true,                 
    numbers=left,                    
    numbersep=3pt,                  
    showspaces=false,                
    showstringspaces=false,
    showtabs=false,                  
    tabsize=2,
    frame=none,                      %
    rulecolor=\color{neuralbackcolour},    %
    lineskip=-0.2em,                 %
    language=Python
}

\newcommand{\jasonhidden}[1]{}

\title{QualityFlow: An Agentic Workflow for Program Synthesis Controlled by LLM Quality Checks}

\author{Yaojie Hu \\
  Iowa State University \\
  \texttt{jhu@iastate.edu} \\\And
  Qiang Zhou \\
  Amazon Web Services\\
  \texttt{zhouqia@amazon.com} \\\And
  Qihong Chen \\
  University of California, Irvine \\ \AND
  Xiaopeng Li, Linbo Liu, Dejiao Zhang, Amit Kachroo, Talha Oz, Omer Tripp \\
  Amazon Web Services
  }

\begin{document}
\maketitle
\begin{abstract}
We introduce QualityFlow, a dynamic agentic workflow for program synthesis. 
Given the English description of a programming problem and a set of unit tests, the model's goal is to synthesize the correct program that solves the problem and passes the tests. 
QualityFlow includes large language model (LLM) agents resembling a software development team, including code generation, testing, and self-debugging.
We propose the LLM Quality Checker, which explicitly ``imagines'' whether the synthesized programs' execution would conform to the unit tests.
The Quality Checks dynamically control the workflow, including actions to submit the final answer, clarify the problem statement, and revert previous workflow steps.
Our experiments show that the Quality Checker can precisely accept any correct program, mitigate faulty synthesized tests, and prevent potential workflow deviation.
QualityFlow establishes the state-of-the-art results on four program synthesis benchmarks: MBPP, HumanEval, and stricter evaluations from MBPP-EvalPlus and HumanEval-EvalPlus.
\end{abstract}

\section{Introduction}
\input{acl/1.introduction}

\section{Related Work}
\input{acl/5.relatedwork}

\section{Methods}
\input{acl/2.methods}

\section{Evaluation}

\input{acl/3.evaluation}

\section{Discussions and Conclusions}

\input{acl/6.conclusions}

\section{Limitations}
We have investigated the effectiveness of our agentic flow for program synthesis using publicly accessible foundation models from Anthropic Claude family and on public benchmark datasets. 
These datasets present a diverse set of programming problems. 
However, there's a possibility that our results may not generalize to other datasets. 
To address this threat and ensure broader applicability, we evaluated the performance of our flow on multiple benchmarks.

All agents in QualityFlow contributes to its state-of-the-art performances, established by our comprehensive results. QualityFlow, as a whole, also improves over zero-shot synthesis consistently. However, an individual agent, the Test Quality Checker, may sometimes have adverse effect on the overall pass@1 if the LLM is not powerful (Opus). Our experiments have pointed out and studied this limitation. The Test Quality Checker pushes the boundary of the quality check idea to validate tests.

The programs that QualityChecker can generate and quality check are programs with a clear set of unit tests. In applications, the unit tests may not always be available for every program file in a large project due to limited test coverage.

In our experiments, we have developed various tools and scripts to facilitate our experiments, and it is conceivable that they might contain bugs. To mitigate this threat, our code base has gone through rigorous code review process. Additionally, we have carried out thorough validity checks and repeated each experiment several times to confirm consistency. Our code will be released publicly for inspection and reproduction.

We measured accuracy using standard pass@k criteria, widely adopted by the research community. The generated program was deemed correct if it passed all test cases. 
The pass@1 examines functional correctness, and no finite test suites can perfectly cover all edge cases. The pass@1 metric is widely used in the literature, and we believe that is the most reasonable performance metric suitable for this study.

\subsection{Potential risks}

Like all program synthesis methods, QualityFlow could generate incorrect programs. Our Code Quality Checker makes a contribution toward mitigating such risks by detecting incorrect programs with high accuracy, but the detection is not perfect.

\subsection{License of artifacts}
We evaluate QualityFlow's program synthesis performance on MBPP \cite{austin2021program}, MBPP-EvalPlus \cite{liu2024your}, HumanEval \cite{chen2021evaluating}, and HumanEval-EvalPlus benchmarks \cite{liu2024your}. 
All these are public research papers with benchmarks under permissible licenses (e.g. CC).

The usage of these artifacts are consistent with their purpose to evaluate program synthesis methods.

\subsection{Parameters For Packages}
The data artifacts are from Huggingface.

\subsection{Use of AI assistants}
AI assistants are used as grammar and spelling checkers in writing of this paper. AI assistants are used to draft programs and debug programs during software engineering. Authors are responsible for all writing and supplemental materials.

\bibliography{my}

\appendix
\label{sec:appendix}

\input{acl/7.appendix}

\end{document}

%% file: acl/1.introduction.tex
\label{sec:intro}

\begin{figure}[t]
    \centering
    \begin{minipage}{\linewidth}
        \textsc{Problem statement: \newline}
        \footnotesize\texttt{Write a function to remove the first and last occurrence of a given character from the string.\newline}
        \normalsize	\textsc{Visible unit tests:}
        \begin{lstlisting}[style=mypython]
assert remove_Occ("hello","l") == "heo"
assert remove_Occ("abcda","a") == "bcd"
assert remove_Occ("PHP","P") == "H"\end{lstlisting}
    \end{minipage}
        \vskip-10pt
    \caption{An example problem from the MBPP code generation benchmark (No. 11).
    How does the model know that the generated code will pass the unit tests? 
    Existing code generation systems assume this implicitly. 
    }
    \label{fig:mbpp}
\end{figure}

Program synthesis is a long-standing goal of software engineering research \cite{manna1971toward, summers1977methodology}, with a history dating back to the 1940s and 50s \cite{backus1957fortran}.
Recently, large language models (LLMs) for program synthesis have created real-world applications that streamline software development and enhance programmer productivity \cite{codet5, deepseekcoder, lu2021codexglue}.
Outperforming zero-shot synthesis with no prohibitive costs to train the LLMs \cite{kaplan2020scaling, wei2022emergent}, Agentic Workflow has recently been proposed as a way to orchestrate multiple LLM agents that form a team to self-reflect \cite{shinn2024reflexion}, debate \cite{khan2024debating}, and solve the problem together in a collaborative style ~\cite{hong2024metagpt}.
Agentic Workflows provide fertile opportunities to incorporate ongoing software engineering research, such as self-debugging \cite{chen2024teaching, agentcoder-self-repair} and automatic test generation \cite{chen2024chatunitest, hanford1970automatic}, for program synthesis systems with all software engineering roles.

We focus on a program synthesis setting that captures the typical software engineering process: given the natural language description (documentation) together with some visible unit tests, the model's goal is to generate a program that solves the problem and passes the evaluation tests (possibly the same as the visible tests) (Figure \ref{fig:mbpp}). 
Program synthesis benchmarks today (e.g. MBPP, HumanEval \cite{mbpp,humaneval,apps}) follow this setting and are incorporated into the standard evaluation suites of large language models (Table~\ref{tab:passatk}).
Existing program synthesis methods have the following limitations: 
\begin{itemize}[left=2pt,topsep=0pt,itemsep=1pt,parsep=0pt,partopsep=0pt]
\item \textit{Assumption of visible unit test conformity.} 
Existing methods typically \textit{assume} that the generated code will follow the visible unit tests required by the problem statement (Figure \ref{fig:mbpp}). 
\item \textit{Bottleneck of synthesized test quality.} Execution of synthesized tests can provide feedback for self-debugging \cite{agentcoder-self-repair}, but incorrect tests may raise unwarranted errors and mislead the self-debugger to turn a correctly synthesized programs into an incorrect one.
\item \textit{Deviation of self-debugging trajectory.} Repeated self-debugging steps may fail to improve intermediate synthesis: it could become stuck in a loop, reach a fixed point, or simply degrade with more bugs and mistakes. Existing multi-agent methods are \textit{static} workflows and fail to explore flexible control flows during code generation \cite{agentcoder-self-repair}.
\end{itemize}

In this paper, we aim to address these limitations and introduce QualityFlow, an Agentic Workflow for program synthesis that resembles a software engineering team with quality assurance at every level. 
QualityFlow includes a Code Generator agent that drafts the program, a Test Designer agent that synthesizes unit tests, and a Self-Debugger agent that iteratively debugs the program based on error messages from failed unit tests.
The agent that controls QualityFlow is the Quality Checker, who decides to invoke other agents: if an intermediate program passes the quality check, it is submitted as the final solution; otherwise, a selected agent continues to synthesize the next intermediate program.
Quality checks are predicted by \textit{Imagined Execution}, an LLM chain-of-thought \cite{chain-of-thought}  self-reflection method that emulate the execution output and compares it with output asserted by the unit test. If the outputs are equal, then the quality check predicts to accept, thereby explicitly examining unit test conformity.
Quality checks can mitigate the adverse impact of incorrectly synthesized tests during self-debugging by detecting incorrect programs.
After self-debugging, if the Quality Checker still does not accept the program, it indicates that the self-debugging trajectory likely failed to reach a good result, and a Clarifier agent will explain the problem statement to remove misunderstandings and restart the Code Generator for one more attempt. 
Finally, if the Quality Checker still rejects, it is likely that the workflow trajectory has fatally deviated, and all self-debugging and clarification steps will be reverted. 
QualityFlow effectively addresses limitations above and achieves the state-of-the-art (SOTA) program synthesis performance on all benchmarks that we evaluate on, including MBPP~\cite{mbpp}, HumanEval \cite{humaneval}, and more extensive evaluations from EvalPlus for both \cite{evalplus}.

The contributions of our paper are as follows: 
\begin{itemize}[left=2pt,topsep=0pt,itemsep=1pt,parsep=0pt,partopsep=0pt]
    \item We introduce QualityFlow, an Agentic Workflow with the SOTA program synthesis performances on four benchmarks: MBPP, MBPP-EvalPlus, HumanEval, and HumanEval-EvalPlus.
    \item We introduce Imagined Execution, a self-reflection method tailored for the program synthesis domain where LLMs predict the correctness of synthesized program through emulated execution with Chain-of-Thought reasoning and explicitly checks for unit test conformity.
    \item We introduce Quality Checker, a controller agent that selects correctly produced programs and navigates the workflow, making control flow decisions---including continuing, restarting, or reverting---based on Imagined Execution.
    \item We study the bottleneck of synthesized tests and introduce Test Quality Checker, extending the idea of quality checks to filter out incorrect tests and improve workflow performance.
    \item We introduce Diversified Prompting that uses a diverse set of prompts in parallel to maximize the possibility that a correct solution is produced and accepted by the quality checker.
\end{itemize}

%% file: acl/5.relatedwork.tex
\paragraph{Code LLM}
Large language models (LLMs) have emerged as a powerful tool for various code-related tasks, including program synthesis~\cite{codegen, codegeex, incoder, codellama, mistral, starcoder2}, bug fixing~\cite{Hossain2024}, program testing~\cite{llm-testing}, and fuzzing~\cite{llm-fuzzing}. 
Through extensive pre-training, they recognize patterns, comprehend context, and generate coherent and contextually relevant code snippets. 

\paragraph{Agentic workflow}
Generation AI has moved from zero-shot synthesis to agentic workflow where multiple LLMs collaborate to produce the answer together, and QualityFlow develops agentic workflow for program synthesis, following \cite{hong2024metagpt, agentcoder-self-repair}.

\paragraph{Self-reflection} 
Self-reflection iteratively enhances the quality of responses generated by large language models (LLMs). \citet{huang2022large} employs a pre-trained LLM to generate high-confidence answers, which are subsequently used to fine-tune the same LLM, effectively improving its performance through self-generated solutions. However, the fine-tuning process can be time-consuming and resource-intensive, and recent work \cite{madaan2024self, khan2024debating} instructs the LLM to provide feedback on its own output, thereby enabling self-refinement without the need for additional training data or reinforcement learning.

%% file: acl/2.methods.tex
\begin{figure}[t]
    \centering
\includegraphics[width=\linewidth]{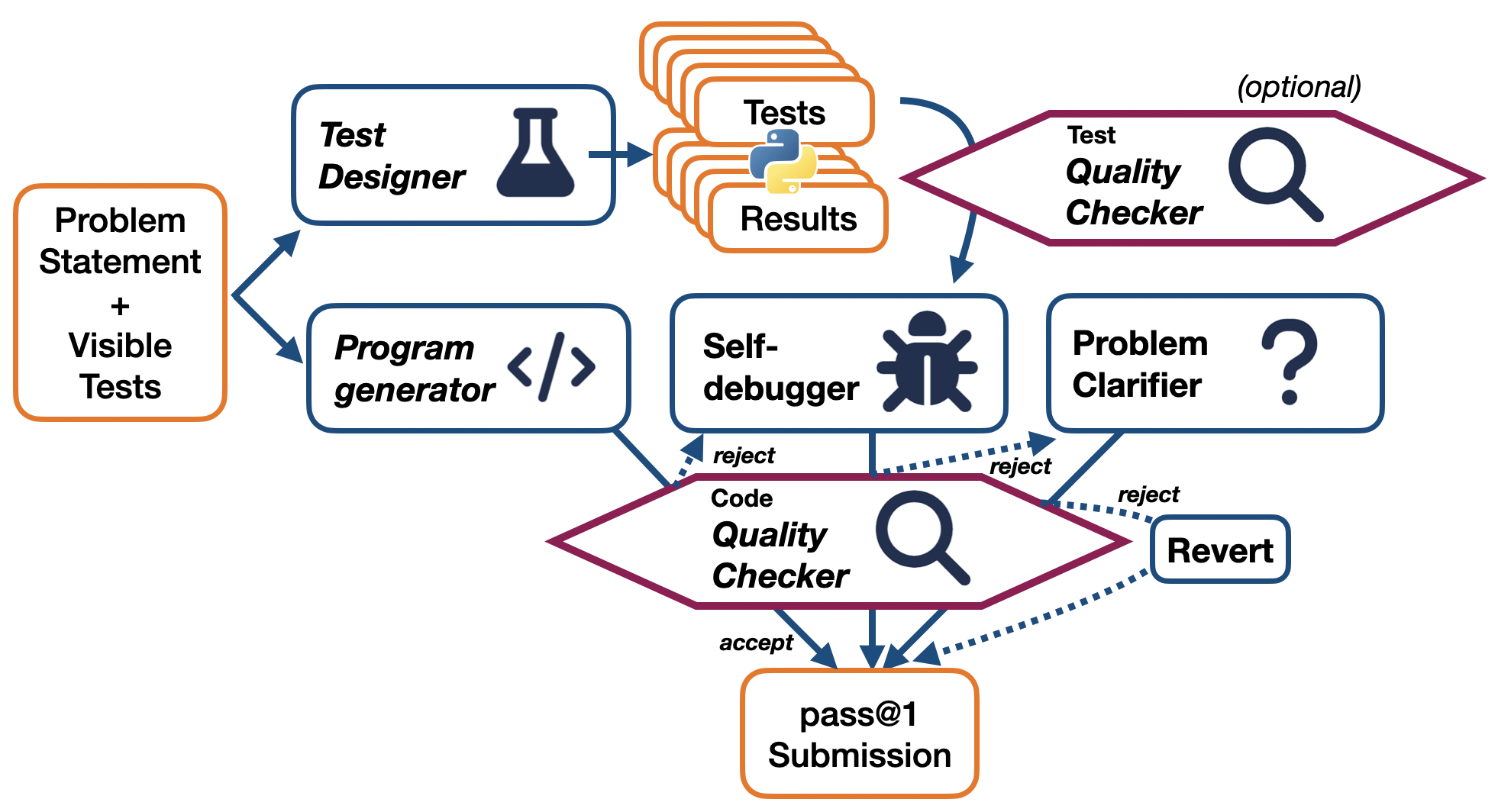}
	\caption{In Quality Flow, the Quality Checker checks the intermediate code at every level.
    If the code passes the quality check, then it is accepted as the final solution; otherwise, next agents perform more processing.
    }
	\label{fig:quality-flow}	
    \vskip -10pt
\end{figure}

\begin{figure}[t]
  \centering
    \resizebox{\linewidth}{!}{%
  \begin{minipage}{1.3\linewidth}
\begin{algorithm}[H]
\caption{QualityFlow}
\label{alg:qf}
\small
\begin{algorithmic}[1]

\State \textbf{Input}: Problem description, visible unit tests
\State \textbf{Output}: A generated program

\State \textbf{Start}

\State GeneratedCode $\gets$ \Call{CodeGenerator}{Problem, VisibleTests} \label{alg:generator}
\State CodeAccept $\gets$ \Call{CodeQualityChecker}{GeneratedCode, Problem, VisibleTests} \label{alg:qc_generator}

\If{CodeAccept}
\State\Return GeneratedCode
\EndIf
\State SynthesizedTests $\gets$ \Call{TestDesigner}{Problem, VisibleTests} \label{alg:test_designer}
\If{Use Test Quality Checker}
    \State FilteredTests $\gets$ \Call{TestQualityChecker}{SynthesizedTests, Problem, VisibleTests} \label{alg:tqc}
\Else
    \State FilteredTests $\gets$ SynthesizedTests
\EndIf

\For{each attempt from 1 to MaxAttempts}
    \State DebuggedCode $\gets$ \Call{SelfDebugger}{GeneratedCode, Problem, FilteredTests} \label{alg:sd}
    \State CodeAccept $\gets$ \Call{CodeQualityChecker}{DebuggedCode, Problem, VisibleTests} \label{alg:qc_sd}
    \If{CodeAccept}
        \State\Return DebuggedCode \label{alg:sd_return}
    \EndIf
\EndFor
\State ClarifiedProblem $\gets$ \Call{ProblemClarifier}{Problem, VisibleTests, DebuggedCode, CodeAccept} \label{alg:clarifier}
\State ClarifiedCode $\gets$ CodeGenerator(ClarifiedProblem, VisibleTests) \label{alg:clarifier_generator}
\State CodeAccept $\gets$ \Call{CodeQualityChecker}{ClarifiedCode, Problem, VisibleTests} \label{alg:qc_clarifier}

\If{CodeAccept}
    \State\Return ClarifiedCode
\Else
    \State\Return GeneratedCode \Comment{Revert} \label{alg:revert}
\EndIf

\State \textbf{End}

\end{algorithmic}
\end{algorithm}
\end{minipage}
}
\vskip -10pt
\end{figure}

QualityFlow is illustrated in Figure \ref{fig:quality-flow}, and a pseudo-code is provided in Algorithm \ref{alg:qf}.

\subsection{Program Generator} The Program Generator agent starts the workflow (line~\ref{alg:generator}). The LLM generates a program based on the problem statement provided, along with a set of visible unit tests that the generated program should pass.
At this stage, no additional contextual information is available; the only input is the original problem statement itself.

\subsection{Test Designer} 
The Test Designer agent (line \ref{alg:test_designer} of Alg. \ref{alg:qf}) synthesizes test cases for the Self-Debugger \cite{chen2024teaching, huang2023agentcoder}.
We instruct the LLM to generate common case unit tests, which leads to higher pass@1 accuracy of the overall workflow than corner case unit tests. 

\subsection{Self-Debugger} 
The Self-Debugger agent (line \ref{alg:sd} of Alg. \ref{alg:qf}) iteratively improves a program through automatic debugging by LLMs with the guidance of synthesized test execution \cite{agentcoder-self-repair, chen2024teaching}.
The Self-Debugger starts with the original problem statement, an intermediate synthesized program, and the synthesized test cases.
The synthesized tests are executed on the program by a Python interpreter to obtain the error messages of the failed test cases and the actual return values of the program. 
Note that running the visible tests can violate the benchmark rules by leaking the ground truth (e.g., MBPP).

Self-debugging is performed through Chain-of-Thought (COT) \cite{chain-of-thought} style reflection, which asks the LLM to analyze step-by-step to reason about why the program fails the test cases and then to offer revisions accordingly, instead of producing a new program directly. 
Self-debugging repeats continues until all tests are passed or the maximum number of epochs is reached, or the Code Quality Checker accepts the program. 

\subsection{Quality Checker}

The (Code) Quality Checker (CQC) examines the quality of the intermediate synthesized programs and navigates the workflow (line \ref{alg:qc_generator}, \ref{alg:qc_sd}, and \ref{alg:qc_clarifier} of Alg. \ref{alg:qf}). 
The Quality Checker makes critical contributions to the state-of-the-art performance of QualityFlow (e.g. on MBPP, 14\% higher pass@1 accuracy, Figure \ref{fig:cqc_removal}). 
The Quality Checker uses \textit{Imagined Execution} to predict code correctness.

\paragraph{Imagined Execution}

In Imagined Execution, the LLM performs Chain-of-Thought reasoning \cite{chain-of-thought} to emulate the execution of a synthesized program given a test input.
The Imagined Execution continues step-by-step and reaches the final return value, and if the result is the same as the test case expects, the Code Quality Checker accepts the program.
If there are multiple tests, the Code Quality Checker verifies all of them, and only if all quality checks pass, the program is considered correct. This strategy leads to high precision (98\% on MBPP, Table \ref{tab:cqc}) and the best pass@1 accuracy for the overall workflow. Recall remains high (98\% on MBPP as well, Table \ref{tab:cqc}), and a slight reduction in recall is not detrimental because the next agents can still produce a correct answer.

\paragraph{Test Quality Checker} 
Quality is a major limitation with the existing test synthesis approach \cite{agentcoder-self-repair, chen2024teaching}, i.e. ``the bottleneck of synthesized test quality''. In experiments, we see that 62\% of LLM-synthesized tests are incorrect (Table \ref{tab:tqc}). Self-debugging may be misled by the erroneous feedback from incorrect tests, and a correctly synthesized program could be changed into an incorrect one. 

The Test Quality Checker (TQC) extends the idea of quality checks to filter out poisonous (incorrect or low-quality) tests.
Given the problem statement and a synthesized test, the Test Quality Checker reasons step-by-step to find the output and compare it with the proclaimed output of the test.
If the two outputs match, the synthesized test is accepted and used in self-debugging; otherwise, it is rejected. 
Note that for the Test Quality Checker, the synthesized program is not an input, because, at this stage, the correctness of the program is in question and cannot be used to judge the synthesized tests. 
In experiments, the Test Quality Checker can identify the incorrect tests with around 80\% recall, improving the overall pass@1 performance by 0.8\% on MBPP (Figure~\ref{fig:tqc}).

\subsection{Problem Clarifier} 
The Problem Clarifier agent explains the problem statement. The problem statement can often be under-specified and misunderstood by programmers or LLMs \cite{mu2023clarifygpt}, leading to incorrect program synthesis during the workflow.
In particular, if the initial understanding by the Code Generator is incorrect, the Self-Debugger could be biased toward the same incorrect interpretation later.
When quality checks rejects all programs from self-debugging, it invokes the Problem Clarifier to explain the problem statement, so the Code Generator has a second chance of generating the correct program. 
The Problem Clarifier not only re-examines the original problem statement but also considers all the contextual information from the test synthesis, test quality check, and self-debugging trajectories. 
The Code Generator has more contextual information this time to understand where the initial misunderstanding of the problem came from and is instructed to produce a different implementation based on the clarified problem statement.

\paragraph{Revert} The re-synthesized program after clarification is given to the Quality Checker for the final check, and if it passes, a submission is made; otherwise, the initial code synthesized by the Code Generator is submitted (line~\ref{alg:revert} in Alg.~\ref{alg:qf}).
Conceptually, knowing that our Quality Checker has high recall (98\% after Clarifier, Table \ref{tab:cqc}), if none of the self-debugging and clarification steps could pass the quality check, it is likely that the workflow has fatally deviated, and the workflow \textit{reverts} to the original synthesis with the least potential for cumulative errors or deviation.

\subsection{Diversified Prompting}
Quality Checker allows us to introduce Diversified Prompting, which uses multiple prompts with slight variations for each agent in the workflow.
In early analysis, we find that different Code Generator prompts can solve different sets of problems with partial overlaps. The pass@1 accuracies of each set are similar, so prompt engineering or prompt selection is not effective.
Inspired by re-ranker models \cite{collins2005discriminative, hu2022fix} that re-orders generated outputs to select the most relevant result post-hoc, we create a diversified set of prompts to increase the chances for a correct solution to be produced and be accepted by the Quality Checker \cite{yao2024tree}. 
For example, instead of a single code generator, we can use multiple (six, for example) code generators in a parallel reasoning structure \cite{yao2024tree}.

In contrast to Self-Consistency that uses the majority vote as the final answer to combat non-determinism and variations in LLM answers \cite{wang2023selfconsistency, chen2024universal, ouyang2024empirical}, Diversified Prompting encourages the diversity of LLM responses to increase the possibility that a correct solution is generated among the candidates, relying on the assumption that the post-hoc re-ranking or quality check will be highly accurate to pick up the correct solution later.
In this sense, Diversified Prompting can be seen as a new post-hoc prompt optimization and selection method \cite{deng-etal-2022-rlprompt, zhang2023automatic}.
Indeed, as our experiments show, our Quality Checker is highly accurate, and, as a result, Diversified Prompting can greatly improve the overall workflow performance (4\% for MBPP, Figure \ref{fig:cqc_removal}).

%% file: acl/3.evaluation.tex
\paragraph{Benchmarks}

\label{sec:benchmark}

We evaluate on MBPP and HumanEval, together with more extensive evaluation on MBPP-EvalPlus and HumanEval-EvalPlus. The four are the competitive program synthesis benchmarks today for LLMs (Table \ref{tab:passatk}). 
We use the standard pass@k metric to assess the correctness of the generated code \cite{kulal2019spoc}. For each programming problem, the model generates $k$ programs, and if any of them pass the unit tests, the problem is considered solved. 
For HumanEval, we experiment with a Relaxed setting where the visible tests are executed on synthesized code to check its quality. MBPP does not allow execution of visible tests because they are the same as evaluation tests.

\paragraph{Models}
We use Claude as the large language model service provider in this paper due to an exclusive research agreement. In this paper, ``Opus'' refers to Claude Opus-3 LLM, and ``Sonnet'' refers to Claude Sonnet-3.5-v2, the latest LLM model available.
Model settings and hyper-parameters can be found in Appendix \ref{appx:hyperparam}.

\subsection{RQ1. Can QualityFlow achieve a new state-of-the-art performance on Python code generation benchmarks?}

\begin{table}[t]
\caption{The pass@1 performance at different workflow steps of QualityFlow, on MBPP. QualityFlow consistently outperforms zero-shot program synthesis with a single-turn LLM prompting across all settings, with absolute improvements in \textbf{bold}.}
\label{tab:steps}
\centering
\resizebox{\linewidth}{!}{

\begin{tabular}{lllrrrrrrrr}
\toprule
Benchmark  & LLM  & Version & Zero-shot &  Prog. Gen. & Debug & Clarifier & Final & $\Delta\uparrow$ \\
       \midrule
\multirow{2}{*}{MBPP}          & Sonnet  & Standard &  81.00 &  86.40 &  93.60 &  93.60 &  94.20 &  \textbf{13.20} \\
                               & Opus    & Standard &  76.00 &  80.80 &  84.60 &  86.20 &  87.00 &  \textbf{11.00} \\
                               \midrule
\multirow{2}{20pt}{MBPP-EvalPlus} & Sonnet  & Standard & 78.31 &  79.37 &  79.10 &  79.63 &  79.89 &  \textbf{1.59} \\
                               & Opus       & Standard & 75.40 &  76.46 &  75.93 &  76.46 &  76.72 &   \textbf{1.32} \\
                               \midrule
\multirow{4}{*}{HumanEval}     & Sonnet  & Standard & 96.34 &  96.34 &  96.95 &  97.56 &  97.56 &   \textbf{1.22} \\
                               & Sonnet  & Relaxed &  96.34 &  97.56 &  98.17 &  98.17 &  98.78 &    \textbf{2.44} \\
                               & Opus    & Standard & 82.32 &  84.76 &  87.20 &  86.59 &  86.59 &   \textbf{4.27} \\
                               & Opus    & Relaxed  & 82.32 &  87.20 &  89.02 &  87.80 &  89.02 &  \textbf{6.71} \\
                               \midrule
\multirow{4}{20pt}{HumanEval-EvalPlus} & Sonnet & Standard & 87.20 &  87.80 &  87.80 &  87.80 &  87.80 & \textbf{0.61} \\
                                       & Sonnet & Relaxed  & 87.20 &  89.02 &  89.63 &  89.63 &  89.63 &   \textbf{2.44} \\
                                       & Opus  & Standard  & 76.22 &  78.05 &  79.88 &  78.66 &  79.27 &   \textbf{3.05} \\
                                       & Opus   & Relaxed  & 76.22 &  79.88 &  81.71 &  81.10 &  81.71 &   \textbf{5.49} \\
\bottomrule
\end{tabular}
}
\end{table}

\begin{table}[t]
\caption{The overall pass@1 performance of QualityFlow compared to the prior reported state-of-the-art results on program synthesis benchmarks. QualityFlow results use Sonnet LLM. Through a flexible agentic workflow with quality checks, QualityFlow has established the state-of-the-art (SOTA) results on all benchmarks (emphasized in bold), with improvements in $\Delta \uparrow$.}
    \label{tab:sota}
    \centering
\resizebox{\linewidth}{!}{
\begin{tabular}{lrrlrr}
\toprule
Benchmark                 & \multicolumn{2}{c}{QualityFlow}                                 & \multicolumn{2}{c}{Prior Published SOTA}        & $\Delta \uparrow$\\
\cmidrule(lr){2-3}\cmidrule(lr){4-5}
                                 & Standard                     & Relaxed                 & Method                               & Result   \\
\midrule               
MBPP                          & \textbf{94.2}   &  N/A   & DeepSeek-Coder-V2-Instruct (\citeyear{deepseekcoder})           & 89.4      & 4.8\\
MBPP-EvalPlus                  & \textbf{79.9}  &  N/A     & DeepSeek-Coder-V2-Instruct (\citeyear{deepseekcoder})           & 76.2      & 3.7\\
HumanEval                    & 97.6             & \textbf{98.8}     & LDB \cite{ldb-self-debug}                                 & 98.2      & 0.6 \\
HumanEval-EvalPlus            & 87.8   & \textbf{89.6}     & Instruct-Turbo \cite{evalplus}               &  86.6     & 3.0  \\
\bottomrule
\end{tabular}
}
\end{table}

We evaluate QualityFlow's program synthesis performance on MBPP \cite{austin2021program}, MBPP-EvalPlus \cite{liu2024your}, HumanEval \cite{chen2021evaluating}, and HumanEval-EvalPlus benchmarks \cite{liu2024your}. 
The step-by-step performance breakdown is in Table \ref{tab:steps}. 

We observe a consistent increase in pass@1 performance as the workflow progresses through the steps. 
As the workflow advances, multiple LLMs collaborate to improve the initial code generated by the Code Generator, leading to an increased pass@1 metric across all settings compared to single-attempt zero-shot synthesis (column $\Delta \uparrow$ in Table \ref{tab:steps}) for all LLM backbones and benchmarks.
QualityFlow can better utilize the programming potential of LLMs without additional training.

\begin{table}[t]
\caption{The pass@k performance on MBPP of QualityFlow compared to previously reported results. QualityFlow achieves a new state-of-the-art performance on MBPP. Different methods vary the number of generated programs submitted for evaluation (parameter $k$ in pass@k), and we use ``-'' to indicate pass@k results that are not reported for the model.
}
\label{tab:passatk}
\centering
\resizebox{\linewidth}{!}{
\begin{tabular}{llcccc}
\toprule
 & \# of prompts &  \multicolumn{4}{c}{pass@} \\          
 \cmidrule{3-6}
&  &  1        & 5          & 10     & 100      \\
\midrule
\textbf{QualityFlow (ours)}            & Multiple       & \textbf{94.2} & \textbf{96.2}        &   -      &    -      \\
AgentCoder (2024) \cite{huang2023agentcoder}            & Multiple       &   -            & <=91.8 &      -   &      -    \\
DeepSeek-Coder-V2-Instruct (2024) \cite{deepseekcoder}       & One   & 89.4    & -   &  -  & - \\ 
Claude Sonnet-3.5 (2024, reproduced)     & One   & 88.7          &     -            &    -     &     -     \\
Claude Opus-3 (2024)                     & One   & 86.4          &     -            &    -     &     -     \\
MapCoder (2024)    \cite{islam-etal-2024-mapcoder}                       & Multiple       & 83.1          &     -            &    -     &    -      \\
Reflexion (2023) \cite{shinn2024reflexion}                        & Multiple       &  77.1 & - & - & - \\ 
Self-Debugging, +Trace (2024) \cite{chen2024teaching}          & Multiple       & 76.4  & - & - & - \\
LDB (2024) \cite{ldb-self-debug} & Multiple & 76.0 & - &- \\
Code Llama (2023)\cite{roziere2023code}                        & One   & 62.4          &   -              & 81.1    & 91.9     \\
Code Llama - Python (2023) \cite{roziere2023code}              & One   & 65.6          &     -            & 81.5    & 91.9     \\
Llama 2 (2023) \cite{touvron2023llama}                           & One   & 45.4          &     -            & 66.2    & 83.1    \\
\bottomrule
\end{tabular}
}
\end{table}

Compared to previously reported state-of-the-art (SOTA) results, QualityFlow has achieved new state-of-the-art performances on all benchmarks, shown in Table \ref{tab:sota}. 
QualityFlow achieved this performance because of careful and accurate quality checks throughout the workflow, which explicitly examines visible unit test conformity, rejects incorrect self-debugging programs misguided by erroneous synthesized tests, and navigate the control flow to rectify deviations of workflow trajectory. Quality checks are analyzed further in Section \ref{sec:cqc_results}.

MBPP provides a suitable setting to show the advantage of the Code Quality Checker and Imagined Execution, as the visible tests cannot be executed, and QualityFlow achieves the state-of-the-art performance of 94.2\% pass@1 accuracy, a 4.8\% increase over the prior SOTA. 
HumanEval allows a \textit{relaxed} setting where QualityFlow uses a Python interpreter that replaces the Code Quality Checker, and QualityFlow achieved 98.8\% pass@1, a 0.6\% improvement over the prior SOTA. 
With a stricter evaluation from HumanEval-EvalPlus, QualityFlow outperformed prior SOTA (86.6\%) with both Imagined Execution Quality Checker (89.6\%) and Python Checker (87.8\%).

We report the pass@k performance of QualityFlow and compare it with prior MBPP results in Table \ref{tab:passatk}.
To report pass@5 accuracy, we choose the last five programs produced from the QualityFlow.
The pass@5 performance of QualityFlow exceeds all results shown in Table \ref{tab:passatk}, including pass@100 results from Llama models, demonstrating QualityFlow's effectiveness and efficiency in settings where multiple generated programs are allowed.

\subsection{RQ2. Can the Quality Checker navigate the workflow and improve accuracy?}

\label{sec:cqc_results}

\begin{table}[t]
\caption{The Code Quality Checker's confusion matrix statistics for different workflow steps. A positive solution passes the evaluation tests. The Code Quality Checker with Imagined Execution is highly accurate.}
\label{tab:cqc}
\resizebox{\linewidth}{!}{
\begin{tabular}{lllrrrrr}
\toprule
          Benchmark                  &        LLM                   & Statistic    & Zero-shot       & Prog. Gen. & Self-Debug & Clarifier  & Final \\
     \midrule
\multirow{10}{*}{MBPP} & \multirow{5}{*}{Sonnet} &      Actual positive &  81.00 &  86.40 &  93.60 &  93.60 &  94.20 \\
                     &                      &             Accuracy      &  93.20 &  95.40 &  96.60 &  96.80 &  96.20 \\
                     &                      &            Precision      &  98.94 &  98.81 &  98.29 &  97.68 &  97.68 \\
                     &                      &               Recall      &  92.59 &  95.83 &  98.08 &  98.93 &  98.30 \\
                     &                      &    \% of final output     &  75.80 &  83.80 &  93.40 &  94.80 &  94.80 \\
                     \cmidrule(lr){2-8}
                     & \multirow{5}{*}{Opus}&      Actual positive  &  76.00 &  80.80 &  84.60 &  86.20 &  87.00 \\
                     &                      &             Accuracy  &  87.20 &  92.60 &  95.00 &  95.00 &  94.20 \\
                     &                      &            Precision  &  98.47 &  97.91 &  95.23 &  95.11 &  95.11 \\
                     &                      &               Recall  &  84.47 &  92.82 &  99.05 &  99.30 &  98.39 \\
                     &                      &    \% of final output &  65.20 &  76.60 &  88.00 &  90.00 &  90.00 \\
                     \midrule
\multirow{10}{50pt}{HumanEval-EvalPlus} & \multirow{5}{*}{Sonnet} &      Actual positive &  95.73 &  96.34 &   96.95 &  97.56 &  97.56 \\
                     &                      &                                   Accuracy &  93.90 &  96.34 &   97.56 &  96.95 &  96.95 \\
                     &                      &                                  Precision &  98.04 &  97.50 &   97.55 &  97.55 &  97.55 \\
                     &                      &                                     Recall &  95.54 &  98.73 &  100.00 &  99.38 &  99.38 \\
                     &                      &                          \% of final output&  93.29 &  97.56 &   99.39 &  99.39 &  99.39 \\
                  \cmidrule(lr){2-8}
                     & \multirow{5}{*}{Opus} &      Actual positive &  89.63 &  92.07 &  95.12 &  94.51 &  95.12 \\
                     &                      &             Accuracy  &  82.93 &  92.07 &  93.29 &  93.29 &  92.68 \\
                     &                      &            Precision  &  96.85 &  96.00 &  95.60 &  95.00 &  95.00 \\
                     &                      &               Recall  &  83.67 &  95.36 &  97.44 &  98.06 &  97.44 \\
                     &                      &    \% of final output &  77.44 &  91.46 &  96.95 &  97.56 &  97.56 \\
         \bottomrule
\end{tabular}
}
\end{table}

\begin{figure}
    \centering
    \includegraphics[width=\linewidth]{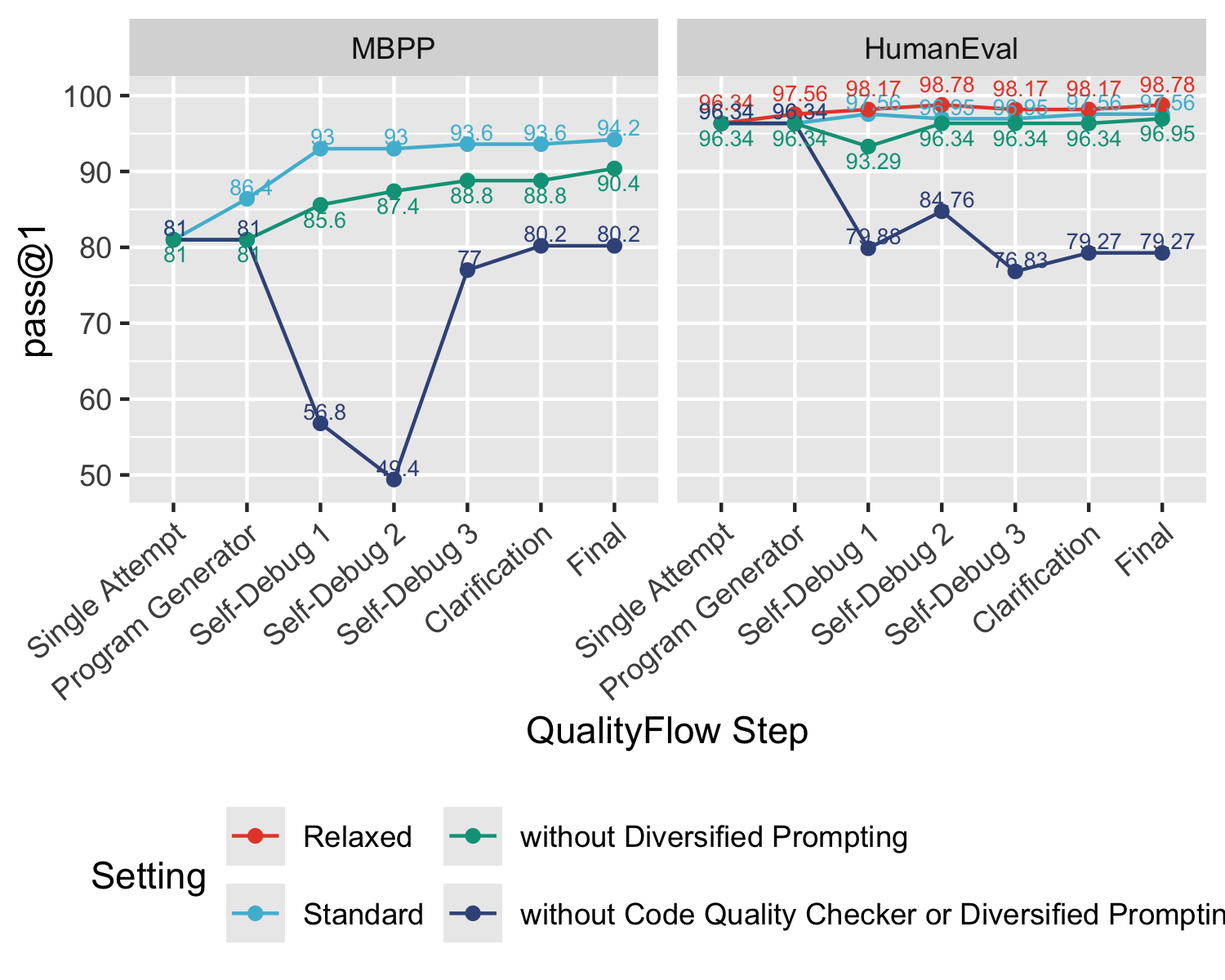}
    \caption{The pass@1 accuracy versus the workflow steps for the standard QualityFlow and the QualityFlow without Code Quality Checker (CQC) with Sonnet LLM. 
    The CQC improves the overall pass@1 performance of the workflow and all workflow steps. 
    }
    \label{fig:cqc_removal}
\end{figure}

Code Quality Checker (CQC) is an LLM agent that uses Imagined Execution, a Chain-of-Thought process, to emulate synthesized programs' execution and predict their correctness by comparing the results with expected outcomes asserted by unit tests. The LLM's algorithmic reasoning ability is essential for accurately emulating program execution and navigating the QualityFlow.

The Code Quality Checker can be independently assessed as a binary classifier. A synthesized program is considered correct if it passes the evaluation tests, and the CQC's objective is to correctly classify them. 
The accuracy and other confusion matrix statistics of Code Quality Checker with Imagined Execution are presented in Table \ref{tab:cqc}.

The precision of the quality check is high (e.g. 98.81\% on the programs produced by Program Generator on MBPP with Sonnet LLM), because we designed for a stricter quality check, where the Imagined Execution results of all visible tests need to pass in order to accept a synthesized program.
This allows the CQC to carefully select correctly synthesized programs and pass the uncertain ones for other agents to solve down the pipeline. 
Shown on the row ``\% of final output'' in Table \ref{tab:cqc}, as the agentic workflow continues, more and more programs are accepted by the Quality Checker.
High accuracy, precision, and recall confirm the effectiveness of Code Quality Checker for classifying the correctness of synthesized programs.

Studying the Code Quality Checker component in isolation is insufficient, as it needs to integrate with the rest of the agentic workflow and improve the overall pass@1 performance. We conducted ablation experiments where the CQC is removed from QualityFlow to measure the effect on the workflow's pass@1 accuracy. 
When the CQC is removed, there will be no termination condition for the workflow, and every problem will go through all workflow steps for pass@1 evaluation. This limitation is seen in existing work \cite{huang2023agentcoder}, and we run QualityFlow without CQC in comparison. Figure \ref{fig:cqc_removal} plots the pass@1 accuracy of all workflow steps with and without the CQC.

Figure \ref{fig:cqc_removal} shows that for the standard QualityFlow with the CQC, the pass@1 accuracy at every step after Code Generator is higher than that without the CQC. The final pass@1 accuracy of the standard setting is 94.2\% on MBPP, higher than 80.2\% without the Code Quality Checker, improving by 14\%.
Without the CQC, all steps in the the workflow will be performed on every program, and correct programs could be changed to incorrect ones during self-debugging guided by incorrectly synthesized tests (Table \ref{tab:tqc}). 
This explains a drastic drop in pass@1 accuracy without the CQC at the Self-Debug epoch 1, from 93\% to 56.8\% on MBPP and from 98.17\% to 79.88\% on HumanEval, reflecting limitations seen in existing work that we discussed previously as the ``bottleneck of synthesized test quality'' and the ``deviation of self-debugging trajectory''.
The CQC can address these two issues by identifying correctly synthesized programs and submit them directly, preventing possible deviation later. When the self-debugger is misled by incorrect tests, the CQC can reject the incorrect programs, invoke the next agent to fix them, restart the workflow, or revert to the initial solution.
The Code Quality Checker plays a effective role in navigating the QualityFlow and can bring clear improvements to the overall performance.

\begin{table}[t]
    \caption{Imagined Execution versus a simple baseline LLM method to directly classify zero-shot program synthesis correctness for Code Quality Checking. The baseline model tends to be overly optimistic and accept incorrect programs, leading to a lower specificity and reduced pass@1 performance. Imagined Execution is critical for QualityFlow's state-of-the-art performance. }
    \centering
\resizebox{\linewidth}{!}{
\begin{tabular}{llrrrrrr}
\toprule
Benchmark & Code Quality Checker & Zero-shot  & Accuracy & Sensitivity & Specificity & Workflow pass@1 \\
\midrule
\multirow{2}{*}{MBPP}      & Yes/No Classifier      &  81.00 &  76.80 &  91.85 &  12.63 & 78.80  \\
                           & Imagined Execution  &  81.00  &  93.20 &  92.59 &  95.79 &  94.20\\
                           \midrule
\multirow{3}{*}{HumanEval} & Yes/No Classifier      &  95.73  &  92.07 &  95.54 &  14.29 & 95.73  \\
                           & Imagined Execution  &  95.73  &  93.29 &  95.54 &  42.86 &  97.56\\
                           & Python checker        &  95.73  & 100.00 & 100.00 & 100.00 & 98.78  \\
\bottomrule
\end{tabular}
}
    \label{tab:imagined}
\end{table}

\paragraph{Imagined Execution versus a simple Yes/No LLM critic.} Without Imagined Execution to emulate program execution and check for unit test conformity, the Code Quality Checker would not be as accurate and would not lead to SOTA workflow performance. We implement a simple baseline that replaces Imagined Execution with an LLM that directly predicts whether the code is correct as a binary classifier. The results are in Table \ref{tab:imagined}.
Our experiments show that the baseline method is overly optimistic that tend to predict that the synthesized programs are correct, shown by the lower specificity than Imagined Execution, along with lower accuracy.
When code quality checking is done with this simple critic, QualityFlow performance can be lower than single-attempt zero-shot program synthesis on MBPP, dropping from 81.0\% to 78.8\%, and equal on HumanEval with 95.73\%.
Indeed, the method for code quality checking is critical for the success of QualityFlow, and using Python interpreter to execute the visible tests to check code creates a perfect classifier and the best QualityFlow performance on HumanEval.

Additional experiments are in the Appendix.

\subsection{RQ3. Can the Test Quality Checker validate LLM-generated tests to improve self-debugging and overall results?}

\begin{table}[t]
\caption{The Test Quality Checker's (TQC's) performance to select incorrect LLM-designed tests. The TQC is performed on the more challenging MBPP and MBPP-EvalPlus problems for which the code generator's program did not pass the Code Quality Checker, and on the triggered synthesized tests for which the code generator's program did not pass. 
The TQC can identify the incorrect tests with good recall and filter out incorrect tests. A positive test is one that is incorrect.}
\label{fig:tqc}
\centering
\resizebox{\linewidth}{!}{
\begin{tabular}{lrrrrrrrr}
\toprule
Benchmark    & \multicolumn{2}{c}{MBPP}  & \multicolumn{2}{c}{MBPP-EvalPlus} & \multicolumn{2}{c}{HumanEval}  & \multicolumn{2}{c}{HumanEval-EvalPlus}  \\
\cmidrule(lr){2-3}\cmidrule(lr){4-5}\cmidrule(lr){6-7}\cmidrule(lr){8-9}
LLM                                 & Sonnet & Opus & Sonnet & Opus & Sonnet & Opus & Sonnet & Opus \\
\midrule
                                    Total programs & 500 & 500 & 378 & 378 & 164 & 164 & 164 & 164 \\
                            Total covered programs & 81 & 117 & 27 & 44 & 4 & 14 & 4 & 14 \\
                                       Total tests & 3992 & 5830 & 1313 & 2188 & 190 & 701 & 190 & 701 \\
                             Total triggered tests & 1963 & 3196 & 502 & 1072 & 39 & 305 & 39 & 305 \\
    Avg. incorrect and triggered tests per problem &  15.67 &  19.63 &  14.91 &  17.12 &   4.25 &   9.71 &   4.25 &   9.64 \\
       Percentage of incorrect and triggered tests &  62.25 &  70.62 &  68.33 &  65.49 &  43.59 &  44.59 &  43.59 &  44.26 \\
       \cmidrule(lr){1-1}\cmidrule(lr){2-3}\cmidrule(lr){4-5}\cmidrule(lr){6-7}\cmidrule(lr){8-9}
                      Accuracy &  55.17 &  70.81 &  61.16 &  70.34 &  46.15 &  84.92 &  46.15 &  84.59 \\
                     Precision &  60.74 &  74.20 &  68.23 &  71.92 &  41.67 &  75.86 &  41.67 &  75.29 \\
                        Recall &  79.13 &  89.94 &  80.76 &  89.74 &  58.82 &  97.06 &  58.82 &  97.04 \\
                            F1 &   0.69 &   0.81 &   0.74 &   0.80 &   0.49 &   0.85 &   0.49 &   0.85 \\
        \cmidrule(lr){1-1}\cmidrule(lr){2-3}\cmidrule(lr){4-5}\cmidrule(lr){6-7}\cmidrule(lr){8-9}
                    Total incorrect tests filtered & 967 & 2030 & 277 & 630 & 10 & 132 & 10 & 131 \\
                 Covered programs before filtering & 78 & 115 & 23 & 41 & 4 & 14 & 4 & 14 \\
                  Covered programs after filtering & 44 & 59 & 13 & 21 & 3 & 10 & 3 & 10 \\
 Avg. triggered tests before filtering &  25.17 &  27.79 &  21.83 &  26.15 &   9.75 &  21.79 &   9.75 &  21.79 \\
Avg. incorrect and triggered tests before filtering &  15.67 &  19.63 &  14.91 &  17.12 &   4.25 &   9.71 &   4.25 &   9.64 \\
  Avg. triggered tests after filtering &   8.43 &   7.80 &   7.38 &   9.33 &   5.00 &  13.10 &   5.00 &  13.10 \\
Avg. incorrect and triggered tests after filtering &   5.80 &   3.85 &   5.08 &   3.43 &   2.33 &   0.40 &   2.33 &   0.40 \\
\bottomrule
\end{tabular}

}
\end{table}

\begin{figure}
    \centering
    \includegraphics[width=\linewidth]{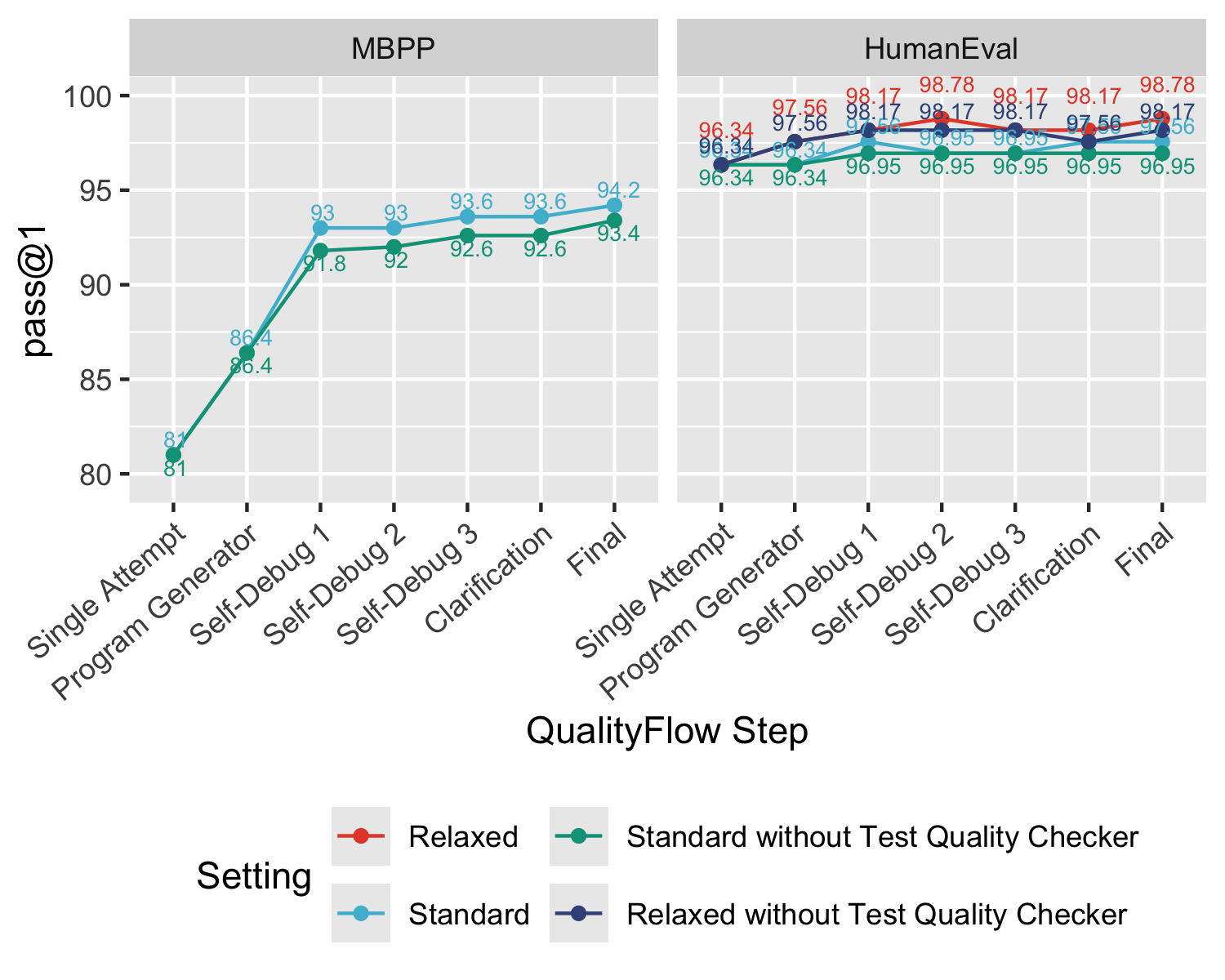}
    \caption{The pass@1 performance of QualityFlow when the Test Quality Checker (TQC) is removed. Test Quality Checker improves the pass@1 accuracy by 1\% on MBPP compared to the standard setting. On HumanEval, removing TQC causes self-debugging performance to drop, but the drop is rectified by Clarifier later.}
    \label{fig:tqc}
\end{figure}

\renewcommand{\arraystretch}{1.05} 
\begin{table}[t]
    \centering
\caption{QualityFlow pass@1 performance when the Test Quality Checker (TQC) is removed from the workflow, compared to the standard QualityFlow. The TQC often improves the overall performance when the better LLM (Sonnet) is used, but it often has negative effect when the weaker LLM is used (Opus), which reflects that the test quality checking problem a more challenging task than code quality checking. 
}
\resizebox{\linewidth}{!}{
\begin{tabular}{lllrrr}
\toprule
LLM & Benchmark & Setting & Standard & without TQC & Delta \\
\midrule
\multirow{6}{*}{Sonnet} & MBPP                                   & Standard &  94.20 &  93.40 &   0.80 \\
                        \cmidrule(lr){2-6}
                        & MBPP-EvalPlus                          & Standard &  79.89 &  79.89 &   0.00 \\
                        \cmidrule(lr){2-6}
                        & \multirow{2}{*}{HumanEval}             & Standard &  97.56 &  96.95 &   0.61 \\
                        &                                        & Relaxed &  98.78 &  98.17 &   0.61 \\
                        \cmidrule(lr){2-6}
                        & \multirow{2}{40pt}{HumanEval-EvalPlus} & Standard &  87.80 &  88.41 &  -0.61 \\
                        &                                        & Relaxed &  89.63 &  89.63 &   0.00 \\
                        \midrule
\multirow{6}{*}{Opus}   & MBPP                                   & Standard &  87.00 &  86.60 &   0.40 \\
                        \cmidrule(lr){2-6}
                        & MBPP-EvalPlus                          & Standard &  76.72 &  77.25 &  -0.53 \\
                        \cmidrule(lr){2-6}
                        & \multirow{2}{*}{HumanEval}             & Standard &  86.59 &  87.20 &  -0.61 \\
                        &                                        & Relaxed &  89.02 &  90.24 &  -1.22 \\
                        \cmidrule(lr){2-6}
                        & \multirow{2}{40pt}{HumanEval-EvalPlus} & Standard &  79.27 &  80.49 &  -1.22 \\
                        &                                        & Relaxed &  81.71 &  82.93 &  -1.22 \\
\bottomrule
\end{tabular}
}
    \label{tab:tqc}
\end{table}
\renewcommand{\arraystretch}{1}

\label{sec:tqc}

We study further on the bottleneck of synthesized test quality and propose Test Quality Checker (TQC) to extend idea of quality checks toward selection of synthesized tests.
The TQC's goal is to predict the correctness of synthesized tests. 
As the synthesized code may not be correct, the TQC predicts the result of a unit test solely based on the problem statement. This contrasts with Imagined Execution in Code Quality Checker, which is provided with both the problem statement and the synthesized code. 
One should reasonably expect a degradation of the emulation execution under such constraints.
Note that while Test Quality Checker has limited effectiveness in addressing the bottleneck of synthesized tests quality, the Code Quality Checker can also detect incorrect programs post-hoc from self-debugging.

The Test Quality Checker can be evaluated independently as a classifier, based on whether it can classify that a synthesized test is correct or incorrect. 
A synthesized test is considered correct if the canonical program, i.e. the standard solution from the benchmark, passes the test. 
The accuracy and confusion matrix statistics are presented in Table \ref{fig:tqc}. The TQC is applied to programs that do not pass the Code Quality Checker, and it assesses synthesized tests that are triggered on the synthesized code with some feedback for self-debugging.
TQC faces a challenging classification problem, because these synthesized programs did not pass the quality check, and any trivial test is easy to pass will not be triggered and assessed by the TQC.

In Table \ref{fig:tqc}, we see 62.25\% synthesized tests for self-debugging are incorrect on MBPP with Sonnet LLM, which empirically proves the bottleneck of synthesized test quality. 
The Test Quality Checker can successfully recall and filter out 79.13\% of the synthesized tests are incorrect, preventing them from misguiding the Self-Debugger. 
We integrate the TQC into QualityFlow and measure the overall pass@1 performance, presented in Table \ref{tab:tqc}. 

Integrating the Test Quality Checker into QualityFlow can conditionally lead to positive improvements on the overall pass@1 accuracy when the better LLM (Sonnet) is used, which contributes to the state-of-the-art performance of QualityFlow for both MBPP and HumanEval by 0.8\% and 0.61\%, respectively. However, when the weaker LLM (Opus) is used, the performance can sometimes drop, reflecting our hypothesis the test quality checking problem is a more challenging task than code quality checking.
As LLMs gain in size and capabilities, the quantitative scaling can lead to emergence of new qualitative abilities \cite{wei2022emergent}, and test quality checking could be an emergent ability enabled only by the latest LLMs such as Sonnet-2.5.
While the TQC contributes to the workflow's state-of-the-art performance, it does not perfectly address the bottleneck of synthesized test quality, as its recall is 80\%, compared to the Code Quality Checker that has 98\% recall. There is room for better test quality checkers to continue to improve self-debugging and program synthesis results in future work.

%% file: acl/6.conclusions.tex
To design an Agentic Workflow tailored for program synthesis, our work identifies three important limitations and opportunities: assumption of visible unit test conformity, bottleneck of synthesized test quality, and deviation of self-debugging trajectory. The three are tackled by one carefully designed controller agent, the Quality Checker, which examines visible test conformity, rejects potentially incorrect programs from self-debugging due to incorrect tests, and restarts or resets potentially deviated trajectories. The Quality Checker uses Imagined Execution and is highly accurate in making these decisions. 
Perhaps for program synthesis, what matters is not to produce the correct solution somewhere in the workflow, but rather to identify and retain them as the final answers. This is the key perspective from the Quality Checker and the reason behind the state-of-the-art results from QualityFlow.

%% file: acl/7.appendix.tex
\section{QualityFlow hyper-parameters}
\label{appx:hyperparam}
For Diversified Prompting, the Program Generator uses 6 diversified prompts to produce varied solutions for each programming problem, some of which are the zero-shot program synthesis prompts seen in prior work \cite{austin2021program, agentcoder-self-repair}. The 6 generated programs form parallel progressions of QualityFlow. For each progression, 3 epochs of self-debugging are performed with a different diversified prompt at each epoch. Afterwards, 3 Clarifier attempts are performed on each of the 6 self-debugging results. 

The temperature is set to 0, except for the Test Designer whose temperature is set to 0.1 to encourage diversity in the tests generated. The Test Designer synthesizes tests in batches of 10 tests per query for 5 rounds, with a maximum of 50 tests per program before post-processing.

\section{DeepSeek integration with QualityFlow}
\label{appx:ablations}

\begin{table}[t]
\caption{We experiment with DeepSeek LLM backbone, and we see that QualityFlow outperforms single-attempt LLM prediction. DeepSeek-v2.5 underperforms Claude Sonnet across all settings, and our work uses Claude Sonnet as the default QualityFlow LLM backbone.}
\label{tab:deepseek}
    \centering
    \resizebox{\linewidth}{!}{
    \begin{tabular}{lllrr}
\toprule
Dataset & Setting& LLM & Single Attempt & QualityFlow pass@1 \\
\midrule
\multirow{3}{*}{MBPP}      & \multirow{3}{*}{Standard} & Deepseek &  79.80 &  87.80 \\
                           &                           & Opus &  80.80 &  87.00 \\
                           &                           & Sonnet &  86.40 &  \textbf{94.20} \\
                                                      \midrule
\multirow{6}{*}{Humaneval} & \multirow{3}{*}{Standard} & Deepseek &  92.68 &  93.90 \\
                           &                           & Opus &  84.76 &  86.59 \\
                           &                           & Sonnet &  96.34 &  \textbf{97.56} \\
                                                      \cmidrule(lr){2-5}
                           & \multirow{3}{*}{Relaxed}  & Deepseek &  92.68 &  96.95 \\
                           &                           & Opus &  87.20 &  89.02 \\
                           &                           & Sonnet &  97.56 &  \textbf{98.78} \\
\bottomrule
    \end{tabular}
}
\end{table}

DeepSeek \cite{deepseekcoder} was the prior SOTA method on MBPP with 89.4\% pass@1 (Table \ref{tab:sota}). It is a LLM and a single-attempt zero-shot program synthesis method. We implemented QualityFlow with the DeepSeek LLM backbone, and results are in Table \ref{tab:deepseek}. The workflow continues to perform better than zero-shot synthesis in all settings. The standard QualityFlow with Sonnet LLM has superior performance than DeepSeek QualityFlow.

\section{LDB reproduction with Sonnet LLM}

\begin{table}[t]
 \caption{We reproduced LDB experiments using the author's publicly released code, using the Sonnet-3.5-v2 LLM backbone, the same one that achieved the new SOTA results on all benchmarks for QualityFlow. On all benchmarks, QualityFlow outperforms LDB.}
    \label{tab:ldb_comparison}
    \centering
        \resizebox{\linewidth}{!}{
    \begin{tabular}{llrlr}
    \toprule
      Benchmark & LLM & Zero-shot & Model & pass@1 accuracy \\
      \midrule
      \multirow{4}{*}{MBPP}        & \multirow{2}{*}{Sonnet-3.5-v1} & \multirow{2}{*}{88.7} & QualityFlow & \textbf{94.2}\\
      &&& LDB & 79.8\\
   \cmidrule(lr){2-5}
                                   & \multirow{2}{*}{Sonnet-3.5-v2} & \multirow{2}{*}{92.2} & QualityFlow & \textbf{95.7} \\
      &&& LDB & 83.0\\
      \midrule
        \multirow{4}{*}{HumanEval} & \multirow{2}{*}{Sonnet-3.5-v1} & \multirow{2}{*}{89.6} & QualityFlow & 93.3 \\
      &&& LDB & \textbf{93.9}\\
   \cmidrule(lr){2-5}
                                   & \multirow{2}{*}{Sonnet-3.5-v2} & \multirow{2}{*}{96.3} & QualityFlow & \textbf{98.8} \\
      &&& LDB & 95.1\\
      \bottomrule
    \end{tabular}
    }
\end{table}

LDB \cite{ldb-self-debug} was the prior SOTA method on HumanEval with 95.1\% pass@1 using the Reflexion LLM backbone \cite{shinn2024reflexion}. To eliminate the confounding effect of different LLMs and compare the program synthesis methods fairly, we reproduced LDB with Sonnet LLM, and results are in Table \ref{tab:ldb_comparison}. LDB also has 95.1\% pass@1, lower than 98.8\% in the authors' original setting (Table \ref{tab:sota}). With the same Sonnet LLM, QualityFlow achieves 98.8\% pass@1 on HumanEval (Table \ref{tab:sota}), proving that the performance advantage comes from our workflow design, not the LLMs used.

\begin{table}[t]
\caption{We conduct ablation experiments that remove Clarifier and Revert separately and measure the overall pass@1 of the entire workflow (Sonnet LLM). Removing each component causes performance drop, and the standard QualityFlow has consistently the best performance.}
\label{tab:removal}
\centering
\resizebox{\linewidth}{!}{
\begin{tabular}{lllr}
\toprule
Dataset & Setting & Ablation & QualityFlow pass@1 \\
\midrule
\multirow{3}{*}{MBPP}      & \multirow{3}{*}{Standard}   & Standard &  \textbf{94.20} \\
                           &                             & Remove Revert &  93.60 \\
                           &                             & Remove Clarifier &  93.80 \\
\midrule
\multirow{6}{*}{HumanEval} & \multirow{3}{*}{Standard}  & Standard &  \textbf{97.56} \\
                           &                            & Remove Clarifier &  \textbf{97.56} \\
                           &                            & Remove Revert &  \textbf{97.56} \\
                         \cmidrule(lr){2-4}
                           & \multirow{3}{*}{Relaxed}   & Standard & \textbf{98.78} \\
                           &                            & Remove Revert &  98.17 \\
                           &                            & Remove Clarifier &  \textbf{98.78} \\
\bottomrule
\end{tabular}
}
\end{table}

\section{Removing the Clarifier or Revert mechanism} 
\label{appx:remove}
To see the effect of the Clarifier and Revert mechanism, we remove the two agents from the workflow and measure the workflow pass@1 performance. On HumanEval, we experiment the standard QualityFlow with Code Quality Checker (CQC) and a relaxed QualityFlow with Python checker, presented in Table \ref{tab:removal}. On MBPP, the standard QualityFlow is used. By removing either Clarifier or Revert, we experiment with a total of six ablation settings, and see reduction of pass@1 across all settings, with reduction ranging from 0.78\% to 2.44\% absolutely.

\section{AgentCoder reproduction}
We have reproduced AgentCoder \cite{huang2023agentcoder} from the author's repository \url{https://github.com/huangd1999/AgentCoder}. We believe that pass@6 evaluation is performed on MBPP, rather than pass@1, for 91.8\% accuracy, inconsistent with the paper's claims.

\section{Token usages}
\begin{table*}[t]
  \caption{Average token count for each component of QualityFlow, as well as the average total tokens per problem.
    As the workflow progresses, if the Code Quality Checker (CQC) accepts an intermediate program, the rest of the steps will be skipped, saving token costs.}
    \label{tab:token}
    \centering
\resizebox{\linewidth}{!}{
\begin{tabular}{llllrrrrrr}
\toprule
Dataset & Code Checker & LLM backbone & Token & Code generator & Test designer & Self debugger & Clarifier & CQC & Total \\
\midrule
\multirow{8}{*}{HumanEval}  & \multirow{4}{*}{Standard CQC}    & \multirow{2}{*}{Opus}   & Input&    486 &   8732 &   5341 &   2452 &   1643 &   4215 \\
                            &                                  &                         & Output&   1087 &   8154 &   2222 &   3136 &   1643 &   4537\\
                            \cmidrule(lr){3-10} 
                            &                                  & \multirow{2}{*}{Sonnet} & Input&    238 &  11295 &   6340 &   4939 &   1060 &   1922 \\
                            &                                  &                         & Output&    594 &   9073 &   2433 &   5356 &   1060 &   2133 \\
                            \cmidrule(lr){2-10} 
                            & \multirow{4}{*}{Python Checker (Relaxed)}  & \multirow{2}{*}{Opus}   & Input&    330 &   8381 &  33124 &   3492 & 0 &   3463 \\
                            &                                  &                         & Output&    784 &   8677 &   2256 &   4583 & 0 &   2265 \\
                             \cmidrule(lr){3-10} 
                            &                                  & \multirow{2}{*}{Sonnet} & Input&    224 &   8970 &  55830 &   4347 & 0 &   2671 \\
                            &                                  &                         & Output&    588 &   9436 &   4475 &   4168 & 0 &   1484 \\
                            \cmidrule(lr){1-10} 
\multirow{4}{*}{MBPP} & \multirow{4}{*}{Standard CQC}         & \multirow{2}{*}{Opus}   & Input&    491 &   6496 &   6144 &   2457 &   3536 &   8722 \\
                            &                                 &                         & Output&   1796 &   9138 &   2692 &   3405 &   3536 &  10019 \\
                                                         \cmidrule(lr){3-10} 
                            &                                 & \multirow{2}{*}{Sonnet} & Input&    255 &   6640 &   8481 &   3334 &   3630 &   7492 \\
                            &                                 &                         & Output&   1122 &   8813 &   3482 &   3601 &   3630 &   7933 \\
\bottomrule
\end{tabular}
}
    \label{tab:my_label}
\end{table*}

\begin{table*}[t]
    \centering
    \caption{Average token count for each component of QualityFlow without Diversified Prompting. The performance is lower, seen in Table \ref{tab:cqc}.}
    \label{tab:token_without_div}
    \resizebox{\linewidth}{!}{
\begin{tabular}{llllrrrrrr}
\toprule
Dataset & Code Checker & LLM backbone & Token & Code generator & Test designer & Self debugger & Clarifier & CQC & Total \\
\midrule
\multirow{8}{*}{HumanEval}  & \multirow{4}{*}{Standard CQC}    & \multirow{2}{*}{Opus} & Input &    141 &  11367 &  11271 &    638 &    662 &   3746 \\
          &                     &                                                    & Output &    124 &  12424 &    583 &    984 &    662 &   2662 \\
           \cmidrule(lr){3-10}
          &                     & \multirow{2}{*}{Sonnet}                             & Input &    141 &  10483 &  10862 &    578 &    501 &   2294 \\
          &                     &                                                     & Output &     71 &   7662 &    876 &    876 &    501 &   1328 \\
                                      \cmidrule(lr){2-10} 
         & \multirow{4}{*}{Python Checker (Relaxed)}  & \multirow{2}{*}{Opus}          & Input &    141 &  10832 &  11410 &    618 & 0 &   2466 \\
         &                   &                                                        & Output &    124 &  11999 &    561 &    901 & 0 &   1528 \\
          \cmidrule(lr){3-10}
         &                   & \multirow{2}{*}{Sonnet}                                 & Input &    141 &  10595 &   9165 &    615 & 0 &   1358 \\
         &                   &                                                         & Output &     71 &   8630 &    797 &    854 & 0 &    661 \\
                                     \cmidrule(lr){1-10} 
\multirow{4}{*}{MBPP} &\multirow{4}{*}{Standard CQC}& \multirow{2}{*}{Opus}           & Input &     85 &   8141 &   3384 &    494 &    762 &   3437 \\
         &                          &                                                  & Output &    401 &  12167 &    762 &    913 &    762 &   4187 \\
                                     \cmidrule(lr){3-10} 
         &                          & \multirow{2}{*}{Sonnet}                          & Input &     85 &   8418 &   3093 &    526 &    858 &   2380 \\
         &                          &                                                  & Output &    463 &   9174 &   1167 &    858 &    858 &   2675 \\
\bottomrule
\end{tabular}
}
\end{table*}

We report the average token count for each component in QualityFlow, as well as the total average token in Table \ref{tab:token}.
The use of Diversified Prompting can increase token usage for each agent, but at the same time, it increase the probability that quality check accepts a program, which can skip the later workflow steps. Without Diversified Prompting, the token usage is fewer, seen in Table \ref{tab:token_without_div}.
Generally, QualityFlow is able to allocate more inference time compute to problems that are more difficult dynamically \cite{jaech2024openai} controlled by Quality Checker.

\clearpage  %
\onecolumn
\section{Examples of Code Quality Checker input and output (Imagined Execution)} 
We provide four examples of Code Quality Checker (CQC) input and output, where the CQC makes two successful and two wrong predictions. In the following input/output conversation, the CQC uses Imagined Execution to find the result of synthesized code execution given a unit test. The examples are from MBPP dataset.

Example 1: A correctly generated code that is correctly accepted.

Input:
\begin{lstlisting}
<function>
```python
def first_Digit(n):
    # Convert number to string and get first character
    return int(str(n)[0])
```
</function>
Think step by step and find the output.
<function_call>
first_Digit(123)
</function_call>
\end{lstlisting}

Output:
\begin{lstlisting}
Let me solve this step by step:
1) The function first_Digit takes a number n as input (in this case, n = 123)
2) Inside the function:
   * str(n) converts 123 to string "123"
   * str(n)[0] gets the first character of the string "123", which is "1"
   * int("1") converts the string "1" back to integer 1
3) Therefore:
   * first_Digit(123) returns 1
The output is: 1
\end{lstlisting}

Input:
\begin{lstlisting}
Given the reasoning above, complete the following test case.
assert first_Digit(123) == ?
Answer in <test_case> ... </test_case> tag.
\end{lstlisting}

Output:
\begin{lstlisting}
assert first_Digit(123) == 1
\end{lstlisting}

The LLM did not follow the instruction to answer with the \texttt{<test\_case>} tag, but the parsing logic handles this scenario and obtains the correct result, which is 1. Correct Imagined Execution like this makes up most of the Code Quality Checker runs, given its high precision (Table \ref{tab:cqc}).

Example 2: A correctly generated code is incorrectly rejected.

Input:
\begin{lstlisting}
<function>
```python
def replace_char(string, old_char, new_char):
    """
    Replace all occurrences of old_char with new_char in the given string.
    
    Args:
        string (str): The input string
        old_char (str): The character to be replaced
        new_char (str): The character to replace with
        
    Returns:
        str: The string with replaced characters
    """
    return string.replace(old_char, new_char)
```
</function>
Think step by step and find the output.
<function_call>
replace_char('polygon', 'y', 'l')
</function_call>
\end{lstlisting}

Output:
\begin{lstlisting}
Let me solve this step by step:
1) We have the input parameters:
   * string = 'polygon'
   * old_char = 'y'
   * new_char = 'l'
2) The function uses the string.replace() method which replaces all occurrences of old_char with new_char
3) Looking at the string 'polygon':
   * It contains one 'y'
   * This 'y' will be replaced with 'l'
4) After replacement:
   * 'polygon' -> 'polgon'
Therefore, the output will be: 'polgon'
\end{lstlisting}

Input:
\begin{lstlisting}
Given the reasoning above, complete the following test case.
assert replace_char('polygon', 'y', 'l') == ?
Answer in <test_case> ... </test_case> tag.
\end{lstlisting}

Output:
\begin{lstlisting}
assert replace_char('polygon', 'y', 'l') == 'polgon'
\end{lstlisting}

LLM made a mistake during Imagined Execution and fails to find the correct result. The correct execution result should be \texttt{assert replace\_char('polygon', 'y', 'l') == 'pollgon'}.

Example 3: An incorrectly generated code is incorrectly accepted. 

Input:
\begin{lstlisting}
<function>
```python
import math
import cmath
def angle_complex(real, imag):
    """
    Calculate the angle (in radians) of a complex number.
    
    :param real: The real part of the complex number
    :param imag: The imaginary part of the complex number
    :return: The angle in radians
    """
    # Create a complex number from the real and imaginary parts
    z = complex(real, imag)
    
    # Use the cmath.phase() function to get the angle
    angle = cmath.phase(z)
    
    # Ensure the angle is in the range [0, 2pi)
    if angle < 0:
        angle += 2 * math.pi
    
    return angle
```
</function>
Think step by step and find the output.
<function_call>
assert math.isclose(angle_complex(0, 1j), ?, rel_tol=0.001)
</function_call>
\end{lstlisting}

Output:
\begin{lstlisting}
Let's think through this step-by-step:
1) The function `angle_complex` takes two parameters: `real` and `imag`.
2) In this case, we're calling `angle_complex(0, 1j)`.
   - The real part is 0
   - The imaginary part is 1j (which is equivalent to i in mathematics)
3) Inside the function, a complex number z is created: z = 0 + 1j
4) The `cmath.phase()` function is used to calculate the angle of this complex number.
5) For the complex number 0 + 1j:
   - This represents a point on the complex plane at (0, 1)
   - This is a point on the positive imaginary axis
   - The angle of this point from the positive real axis is pi/2 radians or 90 degrees
6) The `cmath.phase()` function will return pi/2 for this input.
7) The function then checks if the angle is negative (which it isn't in this case), so no adjustment is needed.
8) The function returns pi/2.
9) In the assertion, we're checking if this result is close to some value (represented by the ? in your question).
10) Given that the expected result is pi/2, we can replace the ? with math.pi / 2.
Therefore, the correct assertion would be:
```python
assert math.isclose(angle_complex(0, 1j), math.pi / 2, rel_tol=0.001)
```
So, the value that should replace the ? is `math.pi / 2`.
\end{lstlisting}

Input:
\begin{lstlisting}
Given the reasoning above, complete the following test case.
assert math.isclose(angle_complex(0, 1j), ?, rel_tol=0.001)
Answer in <test_case> ... </test_case> tag.
\end{lstlisting}

Output:
\begin{lstlisting}
assert math.isclose(angle_complex(0, 1j), math.pi / 2, rel_tol=0.001)
\end{lstlisting}

The canonical test case is \texttt{'assert math.isclose(angle\_complex(0, 1j), 1.5707963267948966, rel\_tol=0.001)'}.
The LLM found the correct answer to the function call, i.e. $\pi/2$, and post-processing code has automatically compared $\pi/2$ to be close to 1.5707963267948966, meaning that the Code Quality Checker passes. But actually, the synthesized program is incorrect: when it executes on the given input, it actually returns 3.141592653589793, i.e. $\pi$, instead.

Example 4: an incorrectly generated code is correctly rejected. 

Input:
\begin{lstlisting}
<function>
```python
def merge_dictionaries_three(dict1, dict2, dict3):
    merged_dict = dict1.copy()  # Start with a copy of the first dictionary
    merged_dict.update(dict2)   # Update with the second dictionary
    merged_dict.update(dict3)   # Update with the third dictionary
    return merged_dict
```
</function>
Think step by step and find the output.
<function_call>
merge_dictionaries_three({'R': 'Red', 'B': 'Black', 'P': 'Pink'}, {'G': 'Green', 'W': 'White'}, {'L': 'lavender', 'B': 'Blue'})
</function_call>
\end{lstlisting}

Output:
\begin{lstlisting}
Let's think through this step-by-step:
1. We start with three dictionaries:
   dict1 = {'R': 'Red', 'B': 'Black', 'P': 'Pink'}
   dict2 = {'G': 'Green', 'W': 'White'}
   dict3 = {'L': 'lavender', 'B': 'Blue'}
2. The function first creates a copy of dict1:
   merged_dict = {'R': 'Red', 'B': 'Black', 'P': 'Pink'}
3. Then it updates merged_dict with dict2:
   merged_dict = {'R': 'Red', 'B': 'Black', 'P': 'Pink', 'G': 'Green', 'W': 'White'}
4. Finally, it updates merged_dict with dict3:
   - It adds the 'L': 'lavender' key-value pair
   - It updates the 'B' key with the new value 'Blue'
5. The final merged_dict looks like this:
   merged_dict = {'R': 'Red', 'B': 'Blue', 'P': 'Pink', 'G': 'Green', 'W': 'White', 'L': 'lavender'}
Therefore, the output of the function call will be:
{'R': 'Red', 'B': 'Blue', 'P': 'Pink', 'G': 'Green', 'W': 'White', 'L': 'lavender'}
Note that the 'B' key, which was present in both dict1 and dict3, has the value 'Blue' in the final dictionary. This is because dict3 was the last to be merged, and its values overwrite any existing keys with the same name.
\end{lstlisting}

Input:
\begin{lstlisting}
Given the reasoning above, complete the following test case.
assert merge_dictionaries_three({'R': 'Red', 'B': 'Black', 'P': 'Pink'}, {'G': 'Green', 'W': 'White'}, {'L': 'lavender', 'B': 'Blue'}) == ?
Answer in <test_case> ... </test_case> tag.
\end{lstlisting}

Output:
\begin{lstlisting}
assert merge_dictionaries_three({'R': 'Red', 'B': 'Black', 'P': 'Pink'}, {'G': 'Green', 'W': 'White'}, {'L': 'lavender', 'B': 'Blue'}) == {'R': 'Red', 'B': 'Blue', 'P': 'Pink', 'G': 'Green', 'W': 'White', 'L': 'lavender'}
\end{lstlisting}

The canonical test case is \texttt{"assert merge\_dictionaries\_three(\{'R': 'Red', 'B': 'Black', 'P': 'Pink'\}, \{'G': 'Green', 'W': 'White'\}, \{'L': 'lavender', 'B': 'Blue'\}) == \{'W': 'White', 'P': 'Pink', 'B': 'Black', 'R': 'Red', 'G': 'Green', 'L': 'lavender'\}"}. 

Executing the code on the test input gives \texttt{\{'R': 'Red', 'B': 'Blue', 'P': 'Pink', 'G': 'Green', 'W': 'White', 'L': 'lavender'\}}, the same as the LLM's reasoning with Imagined Execution. Indeed, the LLM's reasoning is correct: the executed result and the expected result are not the same, where \texttt{B} should be mapped to \texttt{Black} instead of \texttt{Blue}. The synthesized code does not map the \texttt{B} key correctly, thus the synthesized code is wrong, and our Code Quality Checker correctly rejects this code.
Note that the LLM has explained the mapping of \texttt{B} key explicitly, showing the strength of our Imagined Execution method in predicting synthesized code correctness.